\documentstyle[aps,prl]{revtex}

\begin{document}

\draft

\wideabs{
\title{Computing the merger of black-hole binaries: the IBBH problem}
\author{Patrick R. Brady, Jolien D. E. Creighton, and Kip S. Thorne}
\address{Theoretical Astrophysics,
         California Institute of Technology,
         Pasadena, California 91125}
\date{22 April 1998}
\preprint{GRP-498}
\maketitle
\begin{abstract}%
Gravitational radiation arising from the inspiral and merger of binary
black
holes (BBH's) is a promising candidate for detection by kilometer-scale
interferometric gravitational wave observatories.  This paper discusses
a serious obstacle to searches for such radiation and to the
interpretation
of any observed waves: the inability of current computational techniques
to
evolve a BBH through its last $\sim 10$ orbits of inspiral ($\sim 100$
radians of gravitational-wave phase).  A new set of
numerical-relativity techniques is proposed for solving this
``Intermediate
Binary Black Hole'' (IBBH) problem: (i)~numerical evolutions performed in
coordinates co-rotating with the BBH, in which the metric coefficients
evolve
on the long timescale of inspiral, and (ii)~techniques for mathematically
freezing out gravitational degrees of freedom that are not excited by
the waves.
\end{abstract}
}

\narrowtext

\section{Motivation}
\label{sec:motivation}
Among all gravitational wave sources that theorists have considered,
the one most likely to be detected first is the final inspiral and
merger of binary black holes (BBH's) with masses $M_1 \sim M_2
\sim 10$--$20M_\odot$ \cite{bhfirst}.  Detailed analyses of the
evolution of stellar and black-hole
populations~\cite{binary-evolution} predict event rates
as high as $\sim$ one per year in the first LIGO/VIRGO
interferometers (2002--2003) and a thousand per year in enhanced
interferometers for which research and development is currently
under way, but the rates could also be far lower than this.

Optimal search techniques require prior information about the
gravitational waveforms.  The waveforms from the early
binary inspiral phase, when the holes are far apart, are
calculated by a post-Newtonian (PN) expansion.  The merger phase,
beginning at the innermost stable circular orbit, will be calculated
by numerical relativity.  Unfortunately, there is a gap
\cite{Thorne-Finn} between the failure of the PN expansion (which,
for concreteness, we take to occur when its Taylor series makes a
$2\%$ error in the energy loss rate \cite{pnfailure}) and the
beginning of merger.  Filling this gap is called the Intermediate
Binary Black Hole (IBBH) Problem \cite{Thorne-Finn}.

We estimate~\cite{pnfailure} the PN failure point, for calculations at
3PN order [O$(v^6)$ beyond Newtonian gravity and quadrupolar radiation
reaction], to be at the orbital speed $v \equiv (M\Omega)^{1/3}
\simeq 0.3$ (where $M$ is the system's total mass, $\Omega$ is
its orbital angular velocity, and $G=c=1$); there the remaining time to merger,
remaining number of orbits, and remaining number of gravitational-wave
radians are $T \simeq 1200M$, $N_{\rm orbits}
\simeq 8$, and $\Phi \simeq 100$. For 2.5PN calculations, the PN
failure is at $v\simeq0.25$ where $T\simeq5000M$, $N_{\rm orbits}
\simeq 20$, and $\Phi \simeq 250$. For optimal detection of the
waves, the waveform must be accurately modeled in the IBBH
gap~\cite{pnfailure}.  The wave frequency in this gap
is $f = \Omega/\pi \sim (50\;\mbox{to $200$ Hz}) (20M_\odot/M)$,
which is the band of optimal LIGO/VIRGO sensitivity.  This adds
urgency to the IBBH problem.

For numerical simulations of the merger phase, the conventional
approach uses asymptotically inertial coordinates in which the
dynamical timescale, $\tau_{\rm dyn} \sim M$, is set by the task of
moving the holes across the coordinate grid. It is
unlikely that, in the next several years, this approach will be able
to evolve a BBH through the gap for the required $\agt 1200$ dynamical 
time scales.
This motivates exploring alternative procedures for computing the
evolution and waves during the IBBH phase.

One possible method of extending the PN expansion into and through the
IBBH region is to augment it with Pad{\'e}
approximants~\cite{pade-approx}.  Preliminary
results~\cite{pade-approx}, in which PN and exact waveforms in the
test-mass limit are compared, give cause for optimism that the
waveforms' phase evolution can be computed with adequate accuracy via
PN Pad{\'e}-approximants all the way through the IBBH region. However,
there is little hope, via PN Pad{\'e} approximants, to evolve
the binary's internal spacetime geometry in the IBBH
region and thereby provide (i) initial data for numerical relativity's
analysis of the merger, and (ii) a connection between those initial
data and the binary's early inspiral properties (masses, spins, orbit).  For
these crucial issues we must turn elsewhere.

In this paper we explore an alternative strategy~\cite{Thorne-Finn}:
numerical relativity computations performed not in 
asymptotically inertial coordinates (as is normally done), but
instead using spatial coordinates which
co-rotate with the holes' orbital motion and a temporal
slicing which adjusts, as the potential well between the holes
deepens, so as to keep all the metric coefficients as slowly evolving
as possible.  In such coordinates it is reasonable to hope to
achieve a timescale $\tau_{\rm dyn}$ for dynamical evolution of the  
metric coefficients that is of order the timescale $\tau_\ast$
on which radiation reaction drives the holes
together.\footnote{If the holes are spinning with axes inclined to the
orbital angular momentum, then in these coordinates the evolution
timescale may be shorter: $\tau_\ast \sim ($spins' precession
period). For simplicity we shall ignore this possibility, though our
analysis presumably can be adapted to handle it.}  Since the orbital
frequency changes by only a factor of $\sim2$--$3$ through the IBBH
phase, this phase may last only $\sim3$ dynamical timescales in the
co-rotating frame---an enormous reduction from the $\agt1200$
timescales in the asymptotically inertial coordintes of standard
numerical relativity.

Although the metric coefficients' true dynamical timescale 
will be $\tau_\ast$ in these co-rotating coordinates, numerical
approximations may excite spurious gravitational waves with
wavelengths of order the spatial grid size.  Some numerical schemes
will be forced to take time steps shorter than the grid size
in order to suppress these modes and to keep the
numerical evolution stable; cf.\ the Courant condition.  A
good numerical scheme should freeze out these
unphysical modes and stabilize the evolution while using long time steps. 
Correspondingly, a concrete implementation of our strategy must include two
elements: first, a method to choose the lapse and shift so the
coordinates co-rotate with the binary; second, a numerical
scheme that evolves stably with time steps constrained only by
$\tau_\ast$.  Such a scheme differs from that of previous co-rotating
neutron-star-binary calculations~\cite{quasi-ns}, which have
not evolved the gravitational field but instead computed sequences
of equilibria. 

\section{Choosing the lapse and shift}

Numerical relativity is based on a $3+1$ decomposition of the metric:
\begin{equation}
	ds^2 = -\alpha^2 dt^2 + \gamma_{ij} (dx^i + \beta^i dt) (dx^j
+ \beta^j dt) \; . \label{eq:line-element}
\end{equation}
Here $\alpha$ is the lapse function, $\beta^i$ is the shift vector,
and $\gamma_{ij}$ is the intrinsic metric of the 3-dimensional
slices of constant time $t$.  The lapse and shift are specified
freely during the evolution, thereby fixing the spacetime coordinates. 

We propose to construct the initial
IBBH co-rotating coordinates and metric 
from the PN metric near the PN failure point
by adjusting the lapse~$\alpha$ and shift~$\beta_j$ 
so as to make the metric coefficients evolve on the 
inspiral timescale~$\tau_\ast$.  Subsequently $\alpha$ and $\beta_j$ must be
chosen so as
to make the coordinate time derivatives of all the metric
coefficients stay small---or, equivalently in coordinate-independent 
language, to make 
\begin{equation}
	{\cal L}_{\partial_t} \bbox{g} \simeq 0 \;,
\label{e:approx-killing}
\end{equation}
where ${\cal L}_{\partial_t} \bbox{g}$ is the Lie derivative of the spacetime
metric $\bbox{g}$ with respect to the coordinate system's time
generator $\partial_t$.  This is a vague statement, which we make precise by
thinking of the left side of
Eq.~(\ref{e:approx-killing}) as a velocity, constructing
a kinetic energy from this velocity, and choosing a
lapse and shift that minimize this kinetic energy.  We will discuss 
several such action principles for $\alpha$ and $\beta_j$ in the next two 
subsections.

\subsection{The Minimal-Strain Lapse and Shift}

In the spirit of Smarr and York's~\cite{york-smarr} minimal distortion shift, 
we can construct an action principle based on minimizing the Lie derivative 
of the spatial metric $\gamma_{ij}$ rather than the spacetime metric.
Specifically, we presume that the numerical evolution has
proceeded up to some slice of constant time $t$ that has intrinsic
metric $\gamma_{ij}$ and extrinsic curvature $K_{ij}$, and we 
choose the lapse $\alpha$ and shift $\beta_i$ on this slice
so as to minimize the positive definite action
\begin{equation}
        I_1[\alpha,\beta_k] = \int d^3x \! \sqrt{\gamma} \;
	\dot{\gamma_{ij}}
        \gamma^{ik}\gamma^{jl}\dot{\gamma}_{kl} \; .
        \label{e:action-spatial}
\end{equation}
Here $\dot \gamma_{ij} = \partial \gamma_{ij}/ \partial t$ (the Lie
derivative of $\gamma_{ij}$ along 
$\partial_t$) must be expressed in terms of $K_{ij}$, $\alpha$,
$\beta_j$ via the standard relation
\begin{equation}
	\dot \gamma_{ij} = - 2 \alpha K_{ij} + 2D_{(i}\beta_{j)} \; ,
	\label{e:extrinsic-curvature}
\end{equation}
where $D_i$ is the spatial gradient compatible with the
3-metric $\gamma_{ij}$.  By minimizing the resulting action with respect
to variations of $\alpha$ and $\beta_i$, we obtain
four coupled equations:
\begin{mathletters}\label{e:spatial}
\begin{eqnarray}
	K^{ij} [ - 2\alpha K_{ij} + 2D_i\beta_j ] &=& 0 \; ,
	\label{e:spatial-1}\\
	D^j [- 2\alpha K_{ij} + 2D_{(i}\beta_{j)}] &=& 0 \; .
	\label{e:spatial-2}
\end{eqnarray}
\end{mathletters}%
Equation (\ref{e:spatial-1}) is easily solved to give $\alpha$
in terms of $\beta_j$. When that $\alpha$ is inserted
into Eq.~(\ref{e:spatial-2}), the result is a linear, homogeneous
differential equation for $\beta_j$.  If the lapse were not fixed via
Eq.~(\ref{e:spatial-1}) but instead were chosen independent
of $\beta_j$, e.g., via maximal slicing, then the shift
equation (\ref{e:spatial-2}) would reduce to the minimal strain shift
of Smarr and York~\cite{york-smarr}.  We therefore refer
to Eqs.~(\ref{e:spatial}) as {\em minimal strain equations}.

Notice the geometrical nature of the spatial
coordinates carried by this lapse and shift:  The action principle
(\ref{e:action-spatial}) minimizes the rate of change, along
$\partial_t$, of the infinitesimal proper distance between
neighboring points at fixed spatial coordinates.  This
(presumably) will be achieved, in the binary itself, by making the
coordinates co-rotate with the holes, and in the radiation zone by
attaching the spatial coordinates to the wave pattern, i.e., by
(almost) freezing the wave pattern into the spatial coordinate grid.
A direct consequence is that evolution along 
$\partial_t$ is nearly shape and volume preserving.

In the IBBH problem, this approach is not without
shortcomings: there is no guarantee that the minimal strain equations,
which are solved on each spatial slice, will force the lapse and shift
to evolve slowly.  On the other hand, if the initial data are
constructed in coordinates that are close to co-rotating (as
they will be using the known PN metric), and if appropriate
slow-change boundary conditions are enforced on $\beta_i$ near the
holes' apparent horizons and at the outer edge of the coordinate grid,
then it is reasonable to expect $\alpha$ and $\beta_i$ to evolve on
the same slow timescale $\tau_\ast$ as the spatial metric
$\gamma_{ij}$.  This is because $\alpha$ and $\beta_i$ inherit their
dynamics from the time evolution of $\gamma_{ij}$ and $K_{ij}$. 
Note that the minimal strain equations become degenerate
for time-symmetric initial data; such a situation will not arise in
the IBBH problem.
A method of enforcing a variant of Eq.~(\ref{e:spatial-1}) where
$K^{ij}$ is replaced by $\gamma^{ij}$ has been explored by
Balakrishna {\em et al.}~\cite{balakrishna:1996}

The following (far from rigorous) argument makes it
seem likely that this scheme will succeed
for the IBBH problem.  The IBBH spacetime has an ``almost Killing vector
field'' $\bbox{\xi}$, which embodies co-rotation and satisfies
\begin{equation}
	{\cal L}_{\bbox{\xi}} \bbox{g} \equiv \bbox{s} \sim
	\lambda/\tau_*  \; .\label{e:time-derivative}
\end{equation}
Here $\lambda \sim M$ is the length scale over which the spacetime
curvature varies, and $\tau_*\gg \lambda$ is the inspiral timescale,
so $\bbox{s}$ is small.  In terms of the 3+1 spacetime foliation 
being generated by the minimal-strain lapse and shift, we can
decompose $\bbox{\xi}$ into a spatial piece $\bbox{B}$ and
a piece in the direction $\hat{\bbox{n}}$
normal to the surfaces of constant $t$:
$\bbox{\xi} = A \hat{\bbox{n}} + \bbox{B}$,
where $\bbox{B}\cdot\hat{\bbox{n}}=0$ by definition.  We wish to
determine the effectiveness of the minimal strain equations at
attaching the coordinate grid to $\bbox{\xi}$, i.e., at making
$\bbox{\xi} = \partial_t$ or equivalently $A=\alpha$ and $\bbox{B} =
\bbox{\beta}$.  First, we project Eq.~(\ref{e:time-derivative}) into
the spatial slice $\Sigma$ to get
\begin{equation}
	2 D_{(i}B_{j)} - 2 A K_{ij} = s_{ij} \; .
	\label{e:projected-derivative}
\end{equation}
Next, because $\alpha$ and $\beta^i$ satisfy Eqs.~(\ref{e:spatial}),
\begin{equation}
	D^j [- 2 K_{ij} K^{kl} D_k\beta_l / K_{mn}K^{mn}
	+ 2D_{(i}\beta_{j)}] = 0 \; .  \label{e:shifter}
\end{equation}
Finally, substituting $\beta^i=B^i-b^i$ into this equation and using
Eq.~(\ref{e:projected-derivative}) we find that $b^i$, the difference
between the minimal-strain shift and the shift we would like,
satisfies
\begin{eqnarray}
D^j[- 2 K_{ij} && K^{kl} D_k b_l / K_{mn}K^{mn} + 2 D_{(i}b_{j)}]
\nonumber\\
&&= D^j[- K_{ij}(K^{kl} s_{kl} / K_{mn}K^{mn}) + s_{ij} ] \;.
\label{e:difference}
\end{eqnarray}

Assuming (without proof) that the boundary value problem for
Eq.~(\ref{e:shifter}) is well posed, we see that there exists a
solution to Eq.~(\ref{e:difference}) that will scale as $b^i \sim
\lambda/\tau_\ast$; Eq.~(\ref{e:spatial-1}) then implies that $\alpha-A\sim
\lambda/\tau_\ast$.
Therefore, the minimal-strain shift and lapse can make $\partial_t$ equal
to the almost Killing vector field $\bbox{\xi}$ that embodies co-rotation, 
aside from fractional
differences of order $\lambda/\tau_\ast$, as we desired.

Notice that if $\bbox{\xi} \equiv A\hat{\bbox{n}} + \bbox{B}$ is a
Killing vector field on the spacetime then $s_{ij}\equiv 0$,
and $b^i=0$ is a trivial solution to Eqs.~(\ref{e:difference})
corresponding to $\alpha=A$ and $\beta^i=B^i$.

\subsection{Other Choices of Lapse and Shift}
There is much freedom in choosing the
lapse and shift to achieve the goal of slowly evolving metric
coefficients.  Another class of action principles that might 
work
is based on minimizing an integral over spacetime rather than over 3-space as
in Eq.\ (\ref{e:action-spatial}).  Let
$\partial_t = \alpha \hat{\bbox{n}} + \bbox{\beta}$ 
be the vector field to which our coordinates
are tied, and denote the Lie derivative of the 4-metric along
$\partial_t$ by $j_{\mu\nu} = {\cal L}_{{\partial_t}} g_{\mu\nu}$.
Let $\bbox{v}$ be some other vector field independent of
$\partial_t$, and from it construct the tensor
$H^{\mu\nu}_{\bbox{v}} = g^{\mu\nu} + v^\mu v^\nu$.
Then our class of actions is
\begin{equation}
	I_2[\partial_t;\bbox{v}] = \int_{\cal M} (j_{\mu\nu}
	H_{\bbox{v}}^{\mu\rho}
	H_{\bbox{v}}^{\nu\sigma}
	j_{\rho\sigma})\;.
\end{equation}
On varying $\partial_t$,
while holding $\bbox{v}$ and the spacetime metric fixed, we arrive at
\begin{equation}
	\nabla_\nu(H^{\mu\rho}_{\bbox{v}} H^{\nu\sigma}_{\bbox{v}}
	j_{\rho\sigma}) = 0 \; . \label{e:TBC}
\end{equation}
This is a dynamical system of equations for the lapse and shift.
Certain values of $\bbox{v}$ might be considered most natural.  If
$\bbox{v} = \sqrt2 \times$(some unit timelike vector), then
$H_{\bbox{v}}^{\mu\nu}$ is positive definite and there is a
solution of Eqs.~(\ref{e:TBC}) that truly minimizes the action.
If $\bbox{v} = 0$, then Eqs.\  (\ref{e:TBC}) 
are a simple conservation law, but the action is not positive
definite.  It is trivial to show that a spacetime Killing vector field
is a solution to Eqs.\ (\ref{e:TBC}) independent of the choice of $\bbox{v}$, 
and straightforward to extend the analysis of
Eqs.\ (\ref{e:projected-derivative})--(\ref{e:difference}) to show
that for the IBBH problem one of the solutions of
(\ref{e:TBC}) differs from the ``almost Killing vector field'' of Eq.\
(\ref{e:time-derivative}) by an amount that scales as $\lambda/\tau_\ast$.
However, neither here nor for our minimal-strain equations have we
managed to demonstrate that the desired solution for $\partial_t$ is
an attractor; this needs further study.

\section{Numerical Evolution}

To fully solve the IBBH problem will require combining one of our
lapse/shift differential equations with the Einstein equations in some
concrete numerical scheme.  As noted in Sec.\ \ref{sec:motivation}, 
although the binary's metric coefficients should evolve on
the long timescale~$\tau_\ast$ in our proposed co-rotating
coordinate system, there is danger that the time steps will be
driven down to less than the size of the spatial grid by the numerical scheme's
attempt to follow spurious gravitational waves and/or to control numerical
instabilities (the Courant condition). 
To avoid these
pitfalls while taking time steps controlled only by the
inspiral timescale $\tau_\ast$, it will be necessary to stabilize 
the integration scheme and freeze
out the degrees of freedom that are physically present but unphysically
excited.

\subsection{Freezing Out Unwanted Degrees of Freedom}

It is well known that implicit differencing schemes freeze small-scale
structure and produce unconditionally stable evolution.  For this
reason,
we envisage using an implicit scheme to evolve the Einstein equations in
co-rotating coordinates.

While implicit differencing may be sufficient to achieve a stable
evolution on the timescale $\tau_\ast$, we propose an additional
technique that should also help.  The idea is to convert the ADM equations for
$\gamma_{ij}$ and $K_{ij}$ into a parabolic system, thereby removing all
spurious waves while keeping the real ones (which are nearly frozen into 
the co-rotating coordinates).  

The evolution equations written in the usual ADM form are
\begin{mathletters}
\label{e:adm}
\begin{eqnarray}
  \dot \gamma_{ij} &=& -2\alpha K_{ij} + {\cal L}_\beta\gamma_{ij}\;,
  \label{e:admg}\\
  \dot K_{ij} &=& -D_iD_j\alpha
  + \alpha [ {}^3\!R_{ij} + \gamma^{kl}(K_{ij}K_{kl} - 2K_{ik}K_{lj}) ]
  \nonumber\\
  &&\quad + {\cal L}_\beta K_{ij}\;,
  \label{e:admK}
\end{eqnarray}
\end{mathletters}%
where ${}^3\!R_{ij}$ is the Ricci tensor constructed from
$\gamma_{ij}$.  This first-order system can be re-expressed as a second
order system by solving Eq.~(\ref{e:admg}) for the extrinsic curvature
and substituting it into Eq.~(\ref{e:admK}).  Since the fields evolve
on the very long timescale $\tau_\ast$ in the co-rotating
frame, and since $(\partial_t)_{\rm inertial} 
\simeq (\partial_t)_{\rm co-rotating}
+ \Omega \partial_\phi$, the terms with two time derivatives in co-rotating
coordinates will be smaller, by a factor $\sim
1/(\Omega\tau_\ast)$, than
at least some of those with a single time derivative.  Thus, the
double-time-derivative terms can be neglected (or back-differenced, if
desired, so they are treated as sources arising from data on previous
time slices).  In particular, the term involving $\ddot\gamma_{ij}$
can be neglected (or back-differenced).  (Since $\alpha$ and $\beta_i$
are not dynamical fields, we suggest that their time
derivatives also be back-differenced.)  The resulting
parabolic system of equations for $\gamma_{ij}$ can be evolved using
an implicit scheme, which should be stable for large time steps.

\subsection{Initial Data and Boundary Conditions}

To solve the IBBH problem, we must specify suitable initial data and
boundary conditions in addition to formulating an evolution
scheme and a method of choosing the lapse and shift.  One can construct the
initial data, just before the PN failure point, by using matched
asymptotic expansions to join the post-Newtonian exterior metric onto
the metrics of two tidally distorted Kerr black holes, and by then
transforming to co-rotating coordinates~\cite{Kashif}.

The method of Cauchy characteristic matching (e.g., Ref.~\cite{Bishop}) 
seems a promising candidate for constructing
boundary data for evolution of $\gamma_{ij}$ in the co-rotating frame.  Such
matching could conceivably be done around each of the
holes and at an outer boundary in the radiation zone~\cite{Gomez_R:1997}.  
It may be possible to impose outgoing-wave boundary conditions 
as a constraint
on the spatial derivatives of $\gamma_{ij}$ at the outer boundary.  
The shift there is
$\bbox{\beta} \simeq \Omega\partial_\phi$, where $\partial_\phi$ is
the generator of rotations in the orbital plane, and outgoing waves
are constant along $\partial_r + \bbox{\beta}$ (aside from their $1/r$
amplitude falloff), whereas spurious ingoing waves would be constant
along $\partial_r - \bbox{\beta}$.

We also need boundary conditions for the differential equation used
to compute the shift.  Fortunately, these seem easy to
construct.  Since the matching to characteristics would be done on the
history of closed spatial 2-surfaces on the outer boundary and
around each black hole, the derivations of the equations
for the lapse and shift can be repeated for these surfaces.  For
example, the minimal strain equations, obtained by the variations of
an action of the squared velocity of the metric on the 2-surface
[the analog of Eq.~(\ref{e:action-spatial})], are
\begin{mathletters}\label{e:bound_spatial}
\begin{eqnarray}
	k^{ab} [ - 2\alpha k_{ab} + 2{\cal D}_a\beta_b ] &=& 0\; ,
	\label{e:bound_spatial-1} \\
	{\cal D}^a[- 2\alpha k_{ab} + 2{\cal D}_{(a}\beta_{b)}] &=& 0\; ,
	\label{e:bound_spatial-2}
\end{eqnarray}
\end{mathletters}%
where $k_{ab}$ is the extrinsic curvature of the 2-surface embedded
in its history, ${\cal D}_a$ is the covariant derivative operator
compatible with the metric on the 2-surface, and $\beta_a$ is the
3-dimensional shift vector projected into the 2-surface.  Since the
boundary is closed, these equations can be solved to obtain the lapse
function and the tangential components of the shift vector on the
boundary.  The component of the shift that is normal to the boundary
would be set to zero, so the evolution will not attempt to follow
wave crests as they leave the numerical grid.  Among the various
solutions to Eqs.~(\ref{e:bound_spatial}), the one we want will be
that which is closest to the solution on the previous time slice.
This boundary solution, combined with our 3-dimensional differential
equations for the lapse and shift, presumably will produce the desired
$\alpha$ and $\beta_i$, which evolve on the slow time scale
$\tau_\ast$ and co-rotate with the holes.

\section{Worries}

Two conceivably serious difficulties with our approach are: (i) in
our co-rotating reference frame, the almost Killing vector
becomes spacelike beyond the speed-of-light surface, which
might cause problems for the evolution; (ii) when the
second time derivatives are discarded, the resulting evolution might not 
represent the
true evolution of the spacetime.  We doubt, however, that these difficulties 
will
actually arise, since we have seen no sign of them in a  
toy problem that retains the relevant 
features of the IBBH problem.

Our toy problem is a rotating, radiating sphere of scalar
charge in flat spacetime.  We begin in standard spherical polar coordinates
so the line element is given by Eq.~(\ref{eq:line-element}) with
$x^i=\{ r,\theta,\phi\}$, $\gamma_{ij} = \mbox{\rm diag}
[1,r^2,r^2\sin^2\theta]$ and $\beta^i=0$.  The sphere has radius $R$,
angular velocity $\Omega(t)$, moment of inertia $I$ and scalar charge
density $\rho [r,\theta,\phi-\Phi(t)]$ where $\Phi(t)=\int^t dt'
\Omega(t')$.  The scalar field $\Psi$ produced by this charge
distribution satisfies the wave equation
\begin{equation}
	g^{\alpha\beta} \nabla_\alpha\nabla_\beta \Psi = 4 \pi
	\rho \, \Theta (R-r)\; , \label{eq:wave-eqn}
\end{equation}
where $\Theta(R-r)$ is the step function which is unity inside the
sphere and zero outside.  As the sphere rotates, scalar waves are
radiated to infinity, decreasing its angular momentum $I \Omega$ 
according to the radiation-reaction equation
\begin{equation}
	I \frac{\partial \Omega}{\partial t} = 4\pi \int d^3x
	\sqrt{\gamma} \, \bigl[ \rho \,(
	\partial_\phi \Psi) \,\Theta(R-r) \bigr] \; .
\label{e:rad-react}
\end{equation}
For a quadrupolar 
distribution of the scalar charge, this model can be
reduced, via separation of variables, to a (1+1)-dimensional problem,
which we have evolved in the inertial frame by standard 
finite-difference methods for
hyperbolic systems and time steps constrained by the Courant condition.  

We have also evolved this $1+1$ system in the co-rotating frame
$\{r,\theta,\bar\phi=\phi-\Phi(t),t\}$ with $\Phi(t)$ inferred from the
radiation-reaction equation (\ref{e:rad-react}) and {\it not} from some variant
of our minimal-strain equations.  In this co-rotating evolution, we discarded
the (small) second time derivatives of $\Psi$ and applied an 
implicit differencing
scheme to the resulting parabolic system.
We succeeded to
make the time steps as large as the rotational timescale of the
charged sphere (much larger than the Courant condition would allow), and we 
believe that the time steps could be made as
large as the radiation reaction timescale (which is much larger than
the rotational timescale), since the limiting factor was the simple
outer boundary condition that we used.  There were no numerical instabilities.
Moreover, no numerical problems were
encountered at the speed-of-light surface $r=(\Omega\sin\theta)^{-1}$
[and we expected none, since the
transformation to co-rotating coordinates does not change the fact that the
wave equation (\ref{eq:wave-eqn}) is manifestly hyperbolic]. 
There was
good agreement between the results computed in the inertial and
co-rotating frames. Further details will appear elsewhere~\cite{abc}.

Additional evidence that evolution in rotating coordinates need not cause
problems comes from the work of 
Bishop {\it et al.}~\cite{Bishop_N:1997}.  They have used
Schwarzschild spacetime transformed into rigidly rotating coordinates as
a test bed for their numerical evolutions of the Einstein equations, and they
encountered no problems at the speed-of-light surface.  They report that 
their characteristic code ``can
routinely handle such superluminal gauge flows'' at large distances from
the Schwarzschild black hole, even though the coordinate grid moves
slower than the speed of light near the hole.

Based on these results, it seems likely that an implementation of the
methods presented here will allow a numerical evolution of a binary
system through the IBBH phase.  Since a better understanding of this
phase is important---and perhaps critical---for the LIGO/VIRGO
detection of waves from binary black hole systems, and since such systems 
are highly promising candidate sources for LIGO and VIRGO, we hope to
inspire researchers in numerical relativity to address the IBBH
problem via our proposed techniques or others.

\section*{Acknowledgments}

This work was supported in part by NSF grants AST-9417371, AST-9731698 and
PHY-9424337 and NASA grants
NAGW-4268/NAG5-4351 and NAG5-6840.  P.R.B.\ is grateful to the Sherman
Fairchild Foundation for financial support via a Caltech Prize
Fellowship, and J.C.\ is grateful for partial support via a Fellowship
from the Natural Sciences and Engineering Research Council of Canada.
The authors thank Sam Finn for helpful discussions; and 
for helpful critiques of an earlier version of this manuscript, they
thank Miguel Alcubiere, Bernd Bruegmann, Joan Centrella, Carsten Gundlach, 
Richard Matzner, Ed Seidel, Stuart Shapiro, and James York.

\end{document}